\documentstyle[12pt,psfig,draft,lscape,array,amssymb,longtable]{nature-ejb-new}

\def\simlt{\mathrel{\hbox{\rlap{\hbox{\lower4pt\hbox{$\sim$}}}\hbox{$<$}}}}
\def\simgt{\mathrel{\hbox{\rlap{\hbox{\lower4pt\hbox{$\sim$}}}\hbox{$>$}}}}

\def\ale{\mathrel{\hbox{\rlap{\hbox{\lower4pt\hbox{$\sim$}}}\hbox{$<$}}}}
\def\age{\mathrel{\hbox{\rlap{\hbox{\lower4pt\hbox{$\sim$}}}\hbox{$>$}}}}

\def\ra#1#2#3{#1$^{\rm h}$#2$^{\rm m}$#3$^{\rm s}$}
\def\dec#1#2#3{$#1^\circ#2'#3''$}

\newcommand{\swift}{\textit{Swift}}

\def\spose#1{\hbox to 0pt{#1\hss}}
\newcommand\lsim{\mathrel{\spose{\lower 3pt\hbox{$\mathchar"218$}}
     \raise 2.0pt\hbox{$\mathchar"13C$}}}
\newcommand\gsim{\mathrel{\spose{\lower 3pt\hbox{$\mathchar"218$}}
     \raise 2.0pt\hbox{$\mathchar"13E$}}}

\begin{document} 

\title{\Large \bf An extremely luminous X-ray outburst at the
birth of a supernova}

\author{
A.~M.~Soderberg\affiliation[1]{Department of Astrophysical Sciences,
Princeton University, Ivy Lane, Princeton, NJ 08544,
USA}$^,$\affiliation[2]{Carnegie Observatories, 813 Santa Barbara St.,
Pasadena, CA 91101, USA},
E.~Berger\affiliationmark[1]$^,$\affiliationmark[2],
K.~L.~Page\affiliation[3]{Department of Physics and Astronomy, University 
of Leicester, Leicester LE1 7RH, UK},
P.~Schady\affiliation[4]{Mullard Space Sci.~Lab., Univ.~College 
London, Holmbury St.~Mary, Dorking, Surrey RH5 6NT, UK},
J.~Parrent\affiliation[5]{Physics and Astronomy Department Dartmouth 
College, Hanover, NH 03755, USA},
D.~Pooley\affiliation[6]{Astronomy Department, University of Wisconsin, 
475 North Charter Street, Madison, WI 53706, USA},
X.-Y.~Wang\affiliation[7]{Department of Astronomy, Nanjing University, 
Nanjing 210093, China},
E.~O.~Ofek\affiliation[8]{Department of Astronomy, 105-24, California 
Institute of Technology, Pasadena, CA 91125, USA},
A.~Cucchiara\affiliation[9]{Dept.~of Astronomy and Astrophysics, Pennsylvania 
State University, University Park, PA 16802, USA},
A.~Rau\affiliationmark[8],
E.~Waxman\affiliation[10]{Faculty of Physics, Weizmann Institute of 
Science, Rehovot 76100, Israel},
J.~D.~Simon\affiliationmark[8],
D.~C.-J.~Bock\affiliation[11]{Radio Astronomy Laboratory, University of 
California, Berkeley, CA 94720, USA},
P.~A.~Milne\affiliation[12]{Steward Observatory, University of Arizona, 933 
North Cherry Avenue, Tucson, AZ 85721, USA},
M.~J.~Page\affiliationmark[4],
J.~C.Barentine\affiliation[13]{Department of Astronomy, University of Texas at Austin, Austin, TX 78712, USA},
S.~D.~Barthelmy\affiliation[14]{NASA Goddard Space Flight Center, 
Greenbelt, MD 20771, USA},
A.~P.~Beardmore\affiliationmark[3],
M.~F.~Bietenholz\affiliation[15]{Department of Physics and Astronomy, York 
University, Toronto, ON M3J 1P3, Canada}$^,$\affiliation[16]{Hartebeestehoek 
Radio Observatory, PO Box 443, Krugersdorp, 1740, South Africa},
P.~Brown\affiliationmark[9],
A.~Burrows\affiliationmark[1],
D.~N.~Burrows\affiliationmark[9],
G.~Byrngelson\affiliation[17]{Dept.~of Physics and Astronomy, Clemson 
University, Clemson, South Carolina 29634, USA},
S.~B.~Cenko\affiliation[18]{Space Radiation Laboratory 220-47, California 
Institute of Technology, Pasadena, CA 91125, USA},
P.~Chandra\affiliation[19]{Department of Astronomy, University of Virginia, 
P.O. Box 400325, Charlottesville, VA 22904, USA}
J.~R.~Cummings\affiliation[20]{CRESST and NASA Goddard Space Flight Center, 
Greenbelt, MD 20771, USA},
D.~B.~Fox\affiliationmark[9],
A.~Gal-Yam\affiliationmark[10],
N.~Gehrels\affiliationmark[20],
S.~Immler\affiliationmark[20],
M.~Kasliwal\affiliationmark[8],
A.~K.~H.~Kong\affiliation[21]{Institute of Astronomy and Dept. of Physics, National Tsing Hua University, Hsinchu, Taiwan},
H.~A.~Krimm\affiliationmark[20]$^,$\affiliation[22]{Universities Space 
Research Association, 10211 Wincopin Circle, \# 500, Columbia, MD 
21044, USA},
S.~R.~Kulkarni\affiliationmark[8]
T.~J.~Maccarone\affiliation[23]{School of Physics and Astronomy, University of Southampton, Southampton SO17 1BJ, UK},
P.~M\'esz\'aros\affiliationmark[9],
E.~Nakar\affiliation[24]{Theoretical Astrophysics 130-33, California
Institute of Technology, Pasadena, CA 91125, USA},
P.~T.~O'Brien\affiliationmark[3],
R.~A.~Overzier\affiliation[25]{Max-Planck-Institut fur Astrophysik, D-85748 Garching, DE},
M.~de Pasquale\affiliationmark[4],
J.~Racusin\affiliationmark[9],
N.~Rea\affiliationmark[23],
and D.~G.~York\affiliation[26]{Dept.~of Astronomy and Astrophysics, Univ.~of Chicago, 5640 S.~Ellis Avenue, Chicago, IL 60637, USA}
}
\date{\today}{}
\headertitle{A SN at the Time of Explosion}
\mainauthor{Soderberg et al.}

\summary{Massive stars end their short lives in spectacular
explosions, supernovae, that synthesize new elements and drive galaxy
evolution.  Throughout history supernovae were discovered chiefly
through their delayed optical light, preventing observations in the
first moments (hours to days) following the explosion.  As a result,
the progenitors of some supernovae and the events leading up to their
violent demise remain intensely debated.  Here we report the
serendipitous discovery of a supernova at the time of explosion,
marked by an extremely luminous X-ray outburst.  We attribute the
outburst to the break-out of the supernova shock-wave from the
progenitor, and show that the inferred rate of such events agrees with
that of all core-collapse supernovae.  We forecast that future
wide-field X-ray surveys will catch hundreds of supernovae each year in
the act of explosion, and thereby enable crucial neutrino and
gravitational wave detections that may ultimately unravel the
explosion mechanism.}

\maketitle


Stars more massive than about eight times the mass of the Sun meet
their death in cataclysmic explosions termed supernovae (SNe).  These
explosions give birth to the most extreme compact objects -- neutron
stars and black holes -- and enrich their environments with heavy
elements.  It is generally accepted that SNe are triggered when the
stellar core runs out of fuel for nuclear burning and thus collapses
under its own gravity (see Ref.~\pcite{ww86} and references therein).
As the collapsing core rebounds, it generates a shock-wave that
propagates through and explodes the star.

The resulting explosion ejects several solar masses of stellar
material with a mean velocity\cite{fil97} of about $10^4$ km s$^{-1}$,
or a kinetic energy of about $10^{51}$ erg.  Less than a solar mass of
$^{56}$Ni is synthesized in the explosion, but its subsequent
radioactive decay powers\cite{ww86} the luminous optical light
observed to peak 1 to 3 weeks after the explosion.  It is through this
delayed signature that SNe have been discovered both
historically and in modern searches.

While the general picture of core-collapse has been recognized for
many years, the details of the explosion remain unclear and most
SN simulations fail to produce an explosion.  The gaps in our
understanding are due to the absence of detailed observations in the
first days after the explosion, and the related difficulty in detecting the
weak neutrino\cite{abk+89} and gravitational wave signatures of the
explosion.  These signals offer a direct view of the explosion
mechanism but require the discovery of SNe {\it at the time of
explosion}.

In this paper we describe our serendipitous discovery of an extremely
luminous X-ray outburst that marks the birth of a SN of Type
Ibc.  Prompt bursts of X-ray and/or ultraviolet (UV) emission have
been theorized\cite{col74,kc78} to accompany the break-out of the SN
shock-wave through the stellar surface, but their short durations (just
seconds to hours) and the lack of sensitive wide-field X-ray and UV
searches have prevented their discovery until now.

Our detection enables an unprecedented early and detailed view of the
SN, allowing us to infer\cite{wmc07} the radius of the
progenitor star, its mass loss in the final hours prior to the
explosion, and the speed of the shock as it explodes the star.
Drawing on optical, UV, radio, and X-ray observations we show that the
progenitor was compact ($R_*\approx 10^{11}$ cm) and stripped of its
outer Hydrogen envelope by a strong and steady stellar wind.  These
properties are consistent\cite{cgv04} with those of Wolf-Rayet
(WR) stars, the favored\cite{whw02} progenitors of Type Ibc SNe.
 
Wolf-Rayet stars are also argued\cite{mwh01} to give rise to gamma-ray bursts
(GRBs), a related but rare class of explosions characterized
by highly-collimated relativistic jets.  Our observations,
however, indicate an ordinary spherical and non-relativistic explosion
and we firmly rule out a GRB connection.

Most important, the inferred rate of X-ray outbursts indicates that
all core-collapse SNe produce detectable shock break-out emission.
Thus, future wide-field X-ray surveys will uncover hundreds of SNe
each year at the time of explosion, providing the long-awaited
temporal and positional triggers for neutrino and gravitational wave
searches.

\medskip
\noindent
{\bf Discovery of the X-ray Outburst}

\begin{figure}
\centerline{\psfig{file=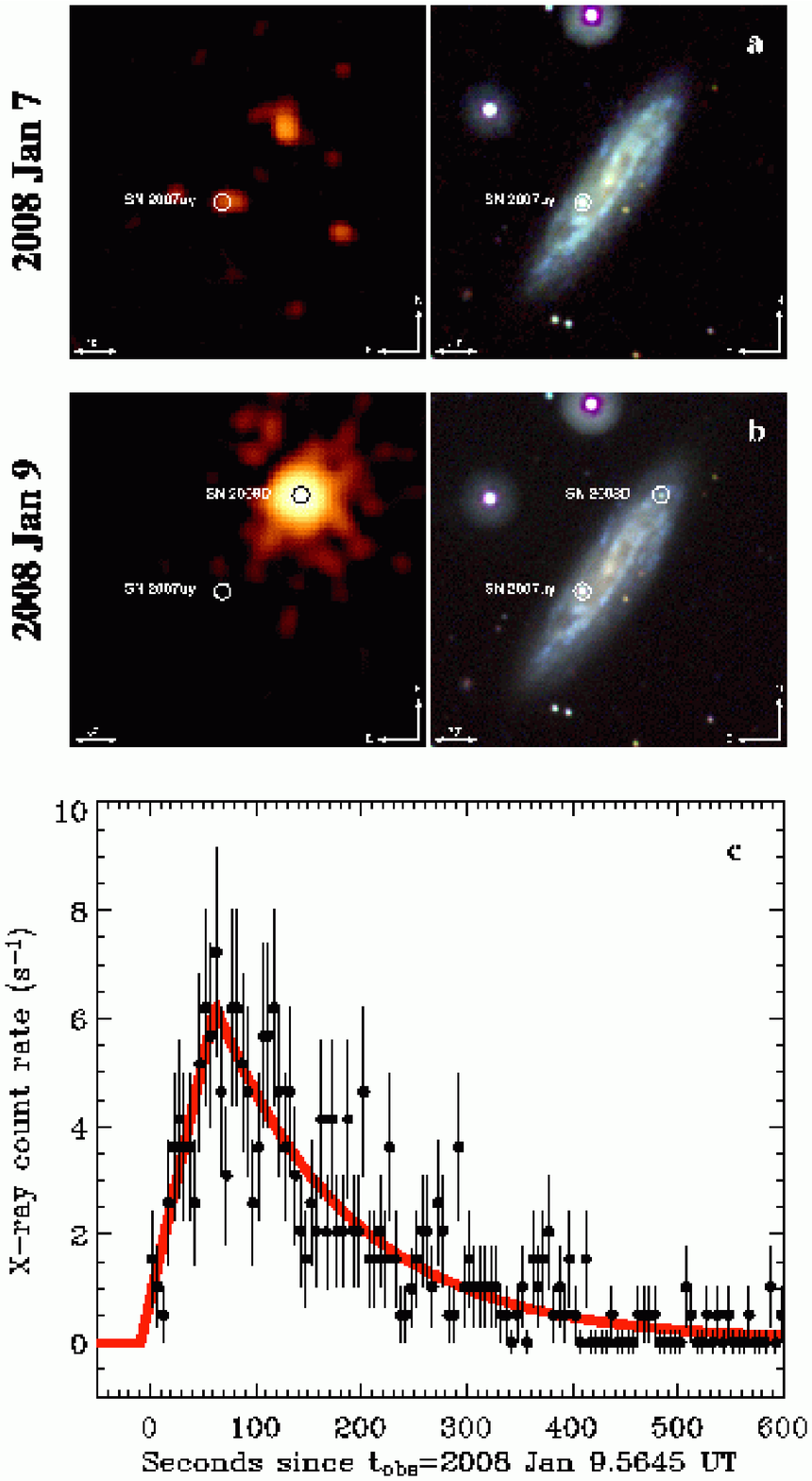,width=5in,angle=0}}
\caption[]{}
\label{fig:disc}
\end{figure}

\noindent
On 2008 Jan 9 at 13:32:49 UT, we serendipitously discovered an extremely
bright X-ray transient during a scheduled \swift\ X-ray Telescope
(XRT) observation of the galaxy NGC\,2770 ($d=27$ Mpc).  Previous XRT
observations of the field just two days earlier revealed no
pre-existing source at this location.  The transient, hereafter
designated as X-ray outburst (XRO) 080109, lasted about 400 s, and was
coincident with one of the galaxy's spiral arms
(Figure~\ref{fig:disc}).  From observations described below
we determine that XRO\,080109 is indeed located in NGC\,2770, and we
thus adopt this association hereafter.

%

The temporal evolution is characterized by a fast rise and exponential
decay, often observed for a variety of X-ray flare phenomena
(Figure~\ref{fig:disc}).  We determine the onset of the X-ray emission
to be $9^{+20}_{-8}$ s before the beginning of the observation,
implying an outburst start time of $t_0 \approx$ Jan 9.564 UT.  The
X-ray spectrum is best fit by a power law [$N(E)\propto E^{-\Gamma}$]
with a photon index of $\Gamma=2.3\pm 0.3$, and a hydrogen column density
of $N_H=6.9^{+1.8}_{-1.5}\times 10^{21}$ cm$^{-2}$, in excess of the
absorption within the Milky Way (see Suppl.~Info.).  The inferred
unabsorbed peak flux is $F_{X,p}\approx 6.9\times 10^{-10}~\rm
erg~cm^{-2}~s^{-1}$ ($0.3-10$ keV).  We also measure significant
spectral softening during the outburst.
 

The XRO was in the field of view of the \swift\ Burst Alert Telescope
(BAT; $15-150$ keV) beginning 30 min before and continuing throughout
the outburst but no $\gamma$-ray counterpart was detected. Thus, the
outburst was not a GRB (see also Suppl.~Info.).  Integrating over the
duration of the outburst, we place a limit on the gamma-ray
fluence of $f_{\gamma}\lsim 8\times 10^{-8}$ erg cm$^{-2}$
($3\sigma$), a factor of three times higher than an extrapolation of
the X-ray spectrum to the BAT energy band.

The total energy of the outburst is thus $E_X\approx
2\times 10^{46}$ erg, at least three orders of magnitude
lower\cite{skb+04} than GRBs.  The peak luminosity is $L_{X,p}\approx
6.1\times 10^{43}$ erg s$^{-1}$, several orders of magnitude larger
than the Eddington luminosity (the maximum luminosity for a
spherically-accreting source) of a solar mass object, outbursts from
Ultra-luminous X-ray sources and Type I X-ray bursts.  In summary, the
properties of XRO\,080109 are distinct from those of all known X-ray
transients.

\medskip
\noindent
{\bf The Birth of a Supernova} 

\noindent Simultaneous observations of the field with the co-aligned
Ultraviolet/Optical Telescope (UVOT) on-board \swift\ showed no
evidence for a contemporaneous counterpart.  However, UVOT
observations just 1.4 hr after the outburst revealed\cite{gcnr110} a
brightening UV/optical counterpart.
Subsequent ground-based optical observations also
uncovered\cite{gcnr110,gcn7160,gcn7163} a coincident source.

\begin{figure}
\centerline{\psfig{file=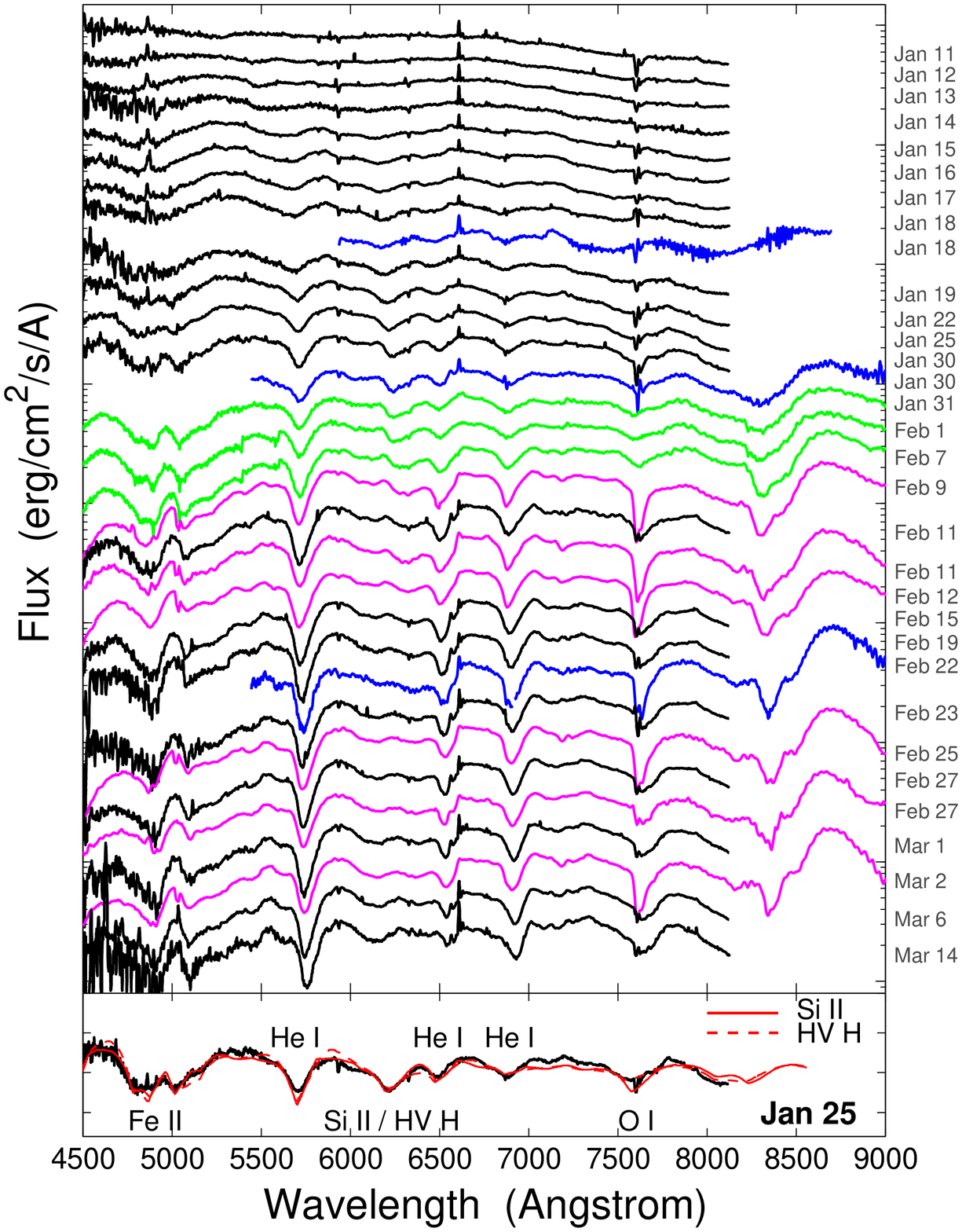,width=7in,angle=0}}
\caption[]{}
\label{fig:optspec} 
\end{figure}

We promptly obtained optical spectroscopy of the counterpart with the
Gemini North 8-m telescope beginning 1.74 d after the outburst
(Figure~\ref{fig:optspec}). The spectrum is characterized by a smooth
continuum with narrow absorption lines of Na I
$\lambda\lambda 5890,5896$ at the redshift of NGC\,2770.  More
importantly, we note broad absorption features near 5200 and 5700
\AA\, and a drop-off beyond 7000 \AA, strongly suggestive of a young
SN.  Subsequent observations confirmed these spectral
characteristics\cite{gcnr110,gcn7169}, and the transient was
classified\cite{gcnr110,cbet1202} as Type Ibc SN\,2008D based on the
lack of hydrogen and weak silicon features.

Thanks to the prompt X-ray discovery, the temporal coverage of our
optical spectra exceeds those of most SNe, rivaling even the
best-studied GRB-associated SNe (GRB-SNe), and SN\,1987A
(Figure~\ref{fig:optspec}).  We see a clear evolution 
from a mostly featureless continuum to broad absorption
lines, and finally to strong absorption features with moderate widths.
Moreover, our spectra reveal the emergence of strong He I features
within a few days of the outburst (see also Ref.~\pcite{cbet1222}).
Thus, SN\,2008D is a He-rich SN Ibc, unlike\cite{pmm+06} GRB-SNe.
Observations at high spectral resolution further reveal significant
host galaxy extinction, with $A_V\approx 1.2-2.5$ mag (see
Suppl.~Info.).

\begin{figure}
\centerline{\psfig{file=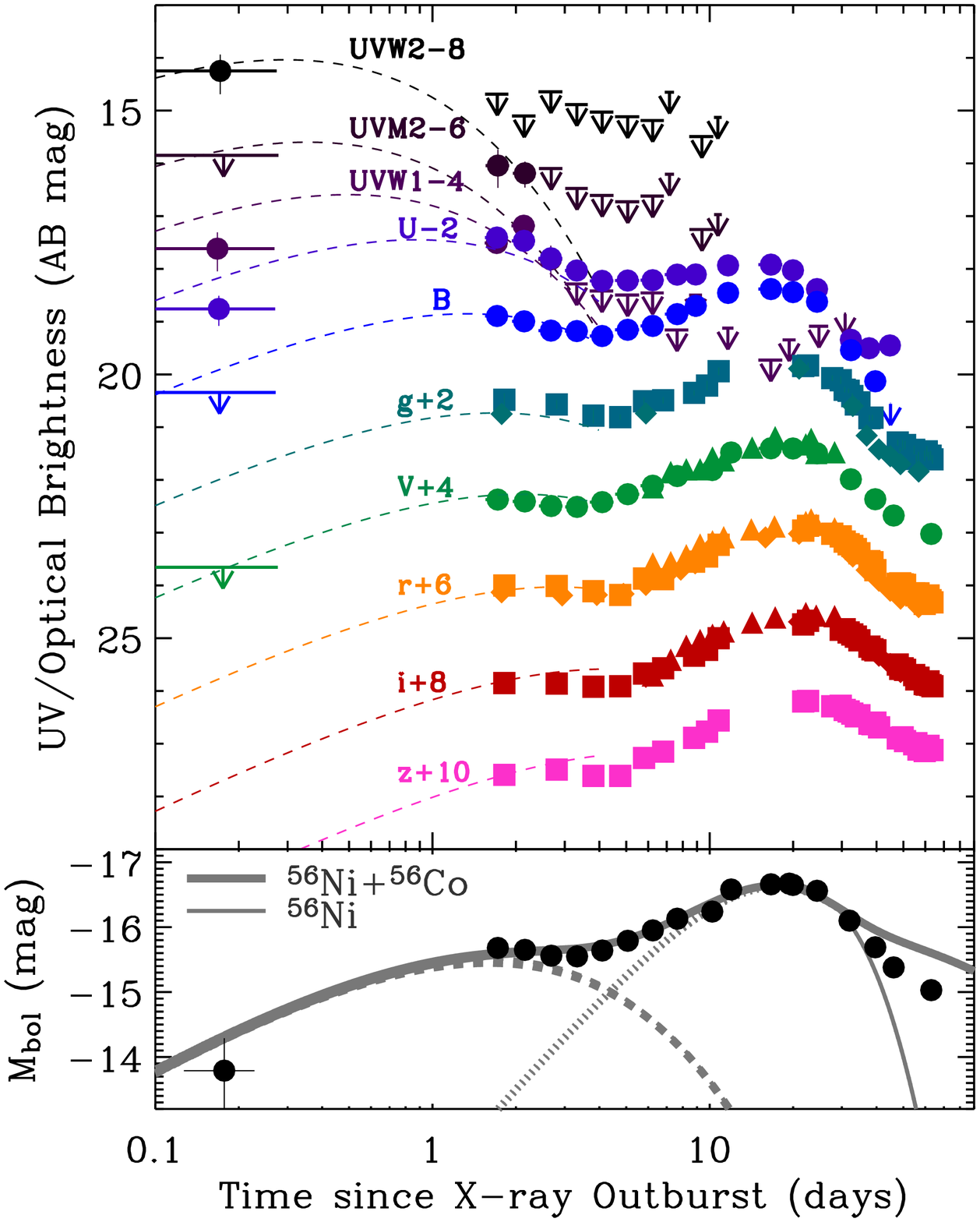,width=6in,angle=0}} 
\caption[]{}
\label{fig:uvopt_lc} 
\end{figure}

The well-sampled UV/optical light curves in ten broadband filters
($2000-10,000$ \AA) exhibit two distinct components
(Figure~\ref{fig:uvopt_lc}).  First, a UV-dominated component that
peaks about a day after the X-ray outburst, and which is similar to
very early observations\cite{cmb+06} of GRB-SN\,2006aj.  The second
component is significantly redder and peaks on a timescale of about 20
days, consistent with observations of all SNe Ibc.  Accounting for an
extinction of $A_V=1.9$ mag (Figure~\ref{fig:uvopt_lc}), the absolute
peak brightness of the second component is $M_V\approx -16.7$ mag, at
the low end of the distribution\cite{skp+06} for SNe Ibc and GRB-SNe.
 
\medskip
\noindent
{\bf A Shock Break-out Origin}

\noindent Since some SNe Ibc harbor GRB jets, we investigate the
possibility that the XRO is produced by a relativistic outflow.  In
this scenario, the X-ray flux and simultaneous upper limits in the
UV/optical require the outflow to be ultra-relativistic with a bulk
Lorentz factor, $\gamma\approx 90$, but its radius to be only
$R\approx 10^{10}$ cm; here $\gamma\equiv (1-\beta^2)^{-1/2}$ and
$\beta\equiv v/c$, where $v$ is the outflow velocity and $c$ is the
speed of light.  However, given the observed duration of the outburst,
we expect\cite{spn98} $R\approx 4\gamma^2ct\approx 10^{17}$ cm,
indicating that the relativistic outflow scenario is not
self-consistent (see Suppl.~Info.~for details).

We are left with a trans- or non-relativistic origin for the outburst
and consider SN shock break-out as a natural scenario.
The break-out is defined by the transition from a radiation-mediated
to a collisional (or collisionless\cite{wl01}) shock as the optical
depth of the outflow decreases to unity.  Such a transition has long been
predicted\cite{col74,kc78} to produce strong, thermal UV/X-ray
emission at the time of explosion.  A non-thermal component at higher
energies may be produced\cite{wlw+07} by multiple scatterings of the photons
between the ejecta and dense circumstellar medium (bulk Comptonization).

We attribute the observed non-thermal outburst to Comptonized emission
from shock break-out indicating that the associated thermal component must
lie below the XRT low energy cutoff, $\sim 0.1$ keV.  With the
reasonable assumption that the energy in the thermal ($E_{\rm th}$)
and Comptonized components is comparable, we constrain\cite{wmc07} the
radius at which shock break-out occurs to $R_{sbo}\gtrsim 7\times
10^{11}(T/0.1\,{\rm keV})^{-4/7}(E_X/2\times 10^{46}\, \rm erg)^{3/7}$
cm.  This is consistent with a simple estimate derived from the rise
time of the outburst, $R_{sbo}=c\delta t\sim 10^{12}$ cm, and larger
than the typical radii of WR stars\cite{mdr89}, $R_*\sim 10^{11}$.  We
therefore attribute the delayed shock break-out to the presence of a
dense stellar wind, similar\cite{wmc07,cmb+06} to the case of
GRB-SN\,2006aj.

The shock velocity at break-out is\cite{wmc07} $(\gamma\beta)\lsim
1.1$ and the outflow is thus trans-relativistic as expected\cite{mm99}
for a compact progenitor.  Using these constraints, the inferred
optical depth of the ejecta to thermal X-rays is $\tau_{\rm ej}\approx
1.5 (E_X/2\times 10^{46}~{\rm erg}) (R_{sbo}/7\times 10^{11}~{\rm
cm})^{-2} (\gamma-1)^{-1}\approx 3$, and Comptonization is thus
efficient, confirming our model.  Equally important, as the ejecta
expand outward the optical depth of the stellar wind decreases and the
spectrum of the Comptonized emission is expected\cite{wlw+07} to
soften, in agreement with the observed trend.


The shock break-out emission traces the wind mass loss rate of the
progenitor, $\dot{M}$, in the
final hours leading up to the explosion.  The inferred density
indicates $\dot{M} \approx 4\pi v_w R_{sbo}/\kappa
\approx 10^{-5}$ M$_{\odot}$ yr$^{-1}$; here $\kappa\approx 0.4~\rm
cm^2~g^{-1}$ is the Thomson opacity for an ionized hydrogen wind and
$v_w\approx 10^3~\rm cm~s^{-1}$ is the typical\cite{cgv04} wind
velocity for WR stars.  The mass loss rate is
consistent\cite{cgv04} with the average values inferred for
Galactic WR stars, and along with the inferred compact stellar radius
and the lack of hydrogen features, leads us to conclude that the
progenitor was a WR star.


\medskip
\noindent
{\bf Two UV/optical Emission Components}

\noindent
The early UV/optical emission ($t\lsim 3$ day)
appears to be a distinct component based on its different temporal
behavior and bluer colors (Figure~\ref{fig:uvopt_lc}).  We
attribute this early emission to cooling of the outer stellar
envelope following the passage of the shock through the star and its
subsequent break-out (marked by the X-ray outburst).  The expected
blackbody radiation is characterized\cite{wmc07} by the photospheric
radius and temperature, which evolve as $R_{\rm ph}\propto t^{0.8}$
and $T_{\rm ph}\propto t^{-0.5}$, and depend on the total ejecta kinetic
energy ($E_K$) and mass ($M_{\rm ej}$), and on the stellar radius
prior to the explosion ($R_*$).

The model light curves provide a good fit to the early UV/optical data
(Figure~\ref{fig:uvopt_lc}). The implied stellar radius is $R_*\approx
7\times 10^{10}$ cm, consistent with that expected\cite{mdr89} for a
Wolf-Rayet progenitor.  Moreover, this value is smaller than the shock
break-out radius, confirming our earlier inference that the break-out
occurs in the extended stellar wind.

The ratio of $E_K$ and $M_{\rm ej}$ also determines the shape of the
main SN light curve (e.g., Ref.~\pcite{vbc+08}), and the mass of
$^{56}$Ni synthesized in the explosion ($M_{\rm Ni}$)
determines\cite{arn82} its peak optical luminosity. To break the
degeneracy between $E_K$ and $M_{\rm ej}$, we measure the photospheric
velocity from the optical spectra at maximum light, $v_{\rm
ph}=0.3(E_{K}/M_{\rm ej})^{1/2}\approx 11,500~\rm km~s^{-1}$,
comparable to that of ordinary SNe Ibc, but somewhat
slower\cite{pmm+06} than GRB-SNe (Figure~\ref{fig:optspec} and
Suppl.~Info).  We find that both light curve components are
self-consistently fit with $E_K\approx (2-4)\times 10^{51}$ erg,
$M_{\rm ej}\approx 3-5$ M$_\odot$, and $M_{\rm Ni}\approx 0.05-0.1$
M$_\odot$ (Figure~\ref{fig:uvopt_lc}).

\medskip
\noindent
{\bf Long-lived X-ray and Radio Emission}

\noindent While UV/optical observations probe the bulk material, radio
and X-ray emission trace fast ejecta.  Our \swift\ follow-up
observations of the XRO revealed fainter X-ray emission several hours
after the explosion with $L_X\approx 2\times 10^{40}$ erg s$^{-1}$
($t\approx 0.2$ d).  This emission exceeds the extrapolation of the
outburst by many orders of magnitude, indicating that it is powered by
a different mechanism.  Using a high angular resolution observation
from the Chandra X-ray Observatory (CXO) on Jan 19.86 UT we detect the
SN with a luminosity, $L_X=(1.0\pm 0.3)\times 10^{39}$ erg s$^{-1}$
($0.3-10$ keV), and further resolve three nearby sources contained
within the $18$ arcsec resolution element of XRT.  Correcting all XRT
observations for these sources, we find that the long-lived X-ray
emission decays steadily as $F_X\propto t^{-0.7}$ (Suppl.~Info.).

Using the Very Large Array (VLA) on Jan 12.54 UT, we further
discovered a new radio source at the position of the SN that was not
present on Jan 7 UT.  Follow-up observations were obtained at
multiple frequencies between 1.4 and 95 GHz using the VLA, the
Combined Array for Research in Millimeter-wave Astronomy (CARMA) and
the Very Long Baseline Array (VLBA).

\begin{figure} 
\centerline{\psfig{file=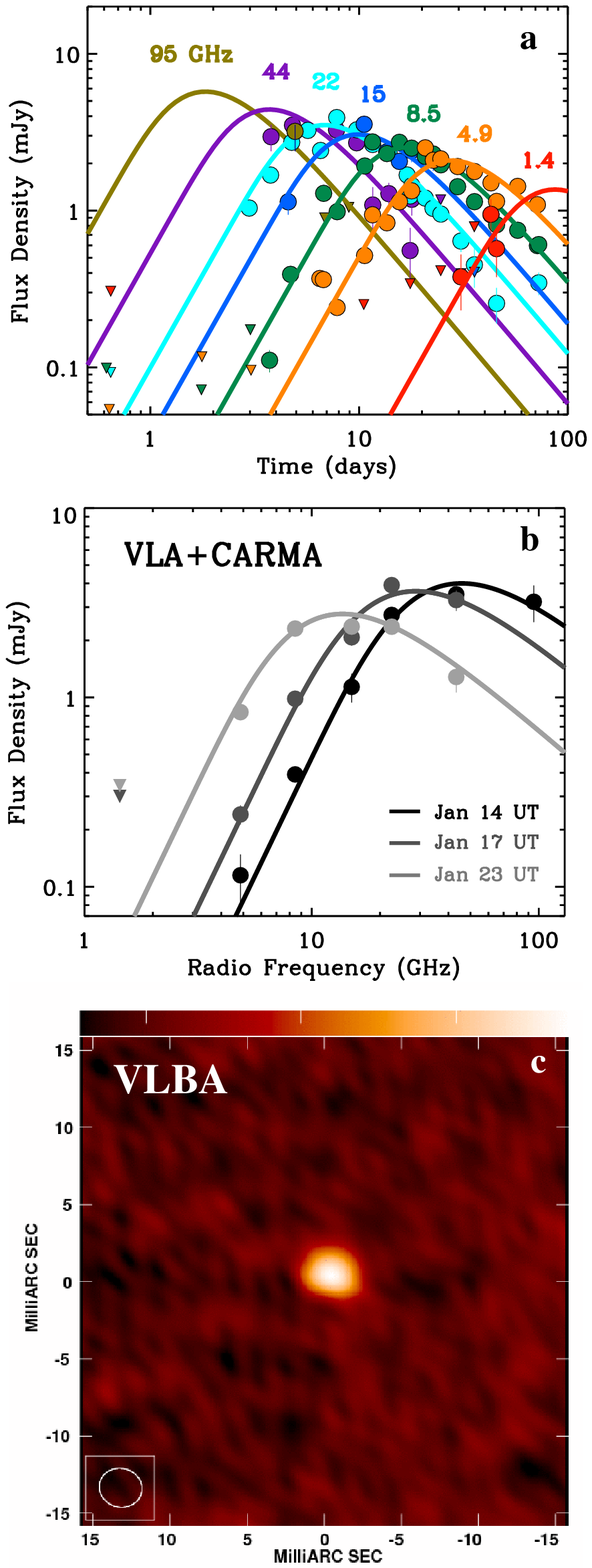,width=7.in,angle=0}}
\caption[]{}
\label{fig:radio} \end{figure}

The broadband radio emission on Jan 14 reveals a spectral peak,
$\nu_p\approx 43$ GHz, with a flux density, $F_{\nu,p}\approx 4$ mJy,
and a low frequency spectrum, $F_{\nu}\propto \nu^{2.5}$.  Subsequent
observations show that $\nu_p$ cascades to lower frequencies,
similar to the evolution observed in other SNe Ibc (e.g.,
Ref.~\pcite{skb+05}).  The passage of $\nu_p$ through each frequency
produces a light curve peak.  We measure $F_{\nu}\propto t^{1.4}$ and
$F_{\nu}\propto t^{-1.2}$ for the light curve rise and decline,
respectively (Figure~\ref{fig:radio}).

We highlight that our X-ray and radio observations of SN\,2008D are
the earliest ever obtained for a normal SN Ibc.  At $t\approx 10$
days, the X-ray and peak radio luminosities are several orders of magnitude
lower\cite{bkf03,fkb+03} than those of GRB afterglows but
comparable\cite{bkf+03,kwp+04} to those of normal SNe Ibc.

\medskip
\noindent
{\bf The Properties of the Fast Ejecta}

\noindent Radio synchrotron emission is produced\cite{che82} by
relativistic electrons accelerated in the SN shock as they gyrate in
the amplified magnetic field.  Self-absorption suppresses the spectrum
below the peak to $F_{\nu}\propto\nu^{2.5}$, in excellent agreement
with our observations.  In this context, we infer\cite{rea94,kfw+98}
the radius of the fast ejecta, using the measured $\nu_p$ and
$L_{\nu,p}$, to be $R\approx 3\times 10^{15}$ cm at $t\approx 5$ d.
The implied mean velocity is $\beta\approx 0.25$, clearly ruling out
relativistic ejecta.

With this conclusion there are two possibilities for the ejecta
dynamics.  First, the SN may be in free expansion, $R\propto t$,
consistent with observations of SNe Ibc (e.g., Ref.~\pcite{skb+05}).
Alternatively, the ejecta may have been relativistic at early time and
then rapidly decelerated, leading to $R\propto t^{2/3}$.  In the
latter scenario the dynamics are governed\cite{wax04} by the
Sedov-Taylor solution.  As discussed in the Suppl.~Info., the temporal
evolution of the radio light curves is clearly inconsistent with the
Sedov-Taylor model, ruling out even early relativistic expansion.

Thus, the radio emission is produced by freely expanding ejecta,
indicative of the broad velocity structure expected\cite{mm99} for
ordinary core-collapse SNe.  The standard formulation\cite{skb+05}
provides an excellent fit to the data (Figure~\ref{fig:radio}) and
indicates that the energy coupled to fast material is $E_{K,R}\approx
10^{48}$ erg, just 0.1\% of the total kinetic energy.  Moreover, the
inferred density profile is $\rho(r)\propto r^{-2}$, as expected for a
steady stellar wind.  The inferred mass loss rate, $\dot{M}\approx
7\times 10^{-6}$ M$_{\odot}$ yr$^{-1}$, is in agreement with our shock
break-out value, indicating stable mass loss in the final $\sim
3$ yr to $\sim 3$ hr of the progenitor's life.  

The radio-emitting electrons also account for the late X-ray emission
through their inverse Compton upscattering of the SN optical photons
(with a luminosity, $L_{\rm opt}$).  The expected\cite{wmc07} X-ray
luminosity is $L_{IC}\approx 3\times 10^{39}\, (E_{K,R}/10^{48}\,{\rm
erg})\, (L_{\rm opt}/10^{42}\,{\rm erg~s^{-1}})\, (t/1\,\rm d)^{-2/3}$
erg s$^{-1}$, in excellent agreement with the XRT and CXO
observations.  We note that the synchrotron contribution in the X-ray
band is lower by at least two orders of magnitude.

Finally, we note that neither the late X-ray emission nor the radio
emission show evidence for a rising component that could be
attributed\cite{snb+06} to an off-axis GRB jet spreading into
our line of sight.  This conclusion is also supported by the
unresolved size of the radio SN from VLBA observations at $t\approx 1$ month, 
$R\simlt 2.4\times 10^{17}$ cm ($3\sigma$), which constrains the
outflow velocity to be $\gamma\beta\simlt 3$.

\medskip
\noindent 
{\bf The Rate of XROs}

\noindent To estimate the rate of XROs we find that the on-sky
effective monitoring time of the XRT from the launch of \swift\
through Jan 2008, including only those exposures longer than 300 s, is
about two years.  Along with the XRT field of view (24 arcmin on a
side), the number density of $L_*$ galaxies ($\phi\approx 0.05$ $L_*$
Mpc$^{-3}$), and the detectability limit of XRT for events like
XRO\,080109 ($d\lsim 200$ Mpc), we infer an XRO rate of $\gsim
10^{-3}$ $L_{*}^{-1}$ yr$^{-1}$ (95\% confidence level); here $L_*$ is
the characteristic luminosity of galaxies\cite{bhb+03}.  This rate is
at least an order of magnitude larger than for
GRBs\cite{sch01,skn+06}.  On the other hand, with a core-collapse SN
rate\cite{cet99} of $10^{-2}$ L$_{*}$ yr$^{-1}$, the probability of
detecting at least one XRO if all such SNe produce an outburst is
about 50\%.

We find a similar agreement with the SN rate using the sensitivity of
the BAT.  The estimated\cite{skn+06} peak photon flux of the outburst
is $0.03$ cm$^{-2}$ s$^{-1}$ ($1-1000$ keV), which for a $10^2$ s
image trigger\cite{ban06} is detectable to about 20 Mpc. The
BAT on-sky monitoring time of 3 years and the 2 ster field of view
thus yield an upper limit on the XRO rate of $\simlt 10^5$ Gpc$^{-3}$
yr$^{-1}$, consistent with the core-collapse SN rate\cite{dsr+04} of
$6\times 10^4$ Gpc$^{-3}$ yr$^{-1}$.

Finally, we note that NGC\,2770 hosted an unusually high rate of three
SNe Ibc in the past 10 years.  However, the galaxy has a typical
luminosity ($0.3$ $L_*$) and a total star formation rate of only
$0.5-1$ M$_\odot$ yr$^{-1}$ (see Suppl.~Info.), two orders of
magnitude lower than the extreme starburst galaxy Arp\,220, which
has\cite{ldt+06} a SN rate of $4\pm 2$ yr$^{-1}$.  The elevated SN
rate in NGC\,2770, with a chance probability of $\sim 10^{-4}$, may
simply be a statistical fluctuation given the sample of $\sim 4\times
10^3$ known SN host galaxies.  Alternatively, it may point to a recent
episode of elevated star formation activity, perhaps triggered by
interaction with the companion galaxy NGC\,2770B at a separation\cite{gcn7186}
of only 22 kpc.

\medskip
\noindent
{\bf Implications for Supernova Progenitors}

\noindent Our observations probe the explosion ejecta over a wide
range in velocity, $\sim 10,000-210,000$ km s$^{-1}$.  Taken together,
the material giving rise to the X-ray outburst, the radio emission,
and the optical light traces an ejecta profile of $E_K\propto
(\gamma\beta)^{-4}$ up to trans-relativistic velocities.  This profile
is in good agreement with theoretical expectations\cite{mm99} for a
standard hydrodynamic spherical explosion of a compact star, but much
steeper\cite{skn+06} than for relativistic GRB-SNe.

On the other hand, we note the similarity between the shock break-out
properties of the He-rich SN\,2008D and the He-poor GRB-SN\,2006aj,
both suggestive of a dense stellar wind around a compact Wolf-Rayet
progenitor.  In the context of SNe Ibc and GRB progenitors, this
provides evidence for continuity (and likely a single progenitor
system) between He-rich and He-poor explosions, perhaps including
GRBs.

Looking forward, our inference that every core-collapse SN is marked
by an XRO places the discovery and study of SNe on the cusp of a
paradigm shift.  An all-sky X-ray satellite with a sensitivity similar
to that of the \swift/XRT will detect and localize several hundred
core-collapse SNe per year, even if they are obscured by dust, {\it at
the time of explosion}.  As we have shown here, this will enable a
clear mapping between the properties of the progenitors and those of
the SNe.  Most important, however, X-ray outbursts will provide an
unprecedented positional and temporal trigger for neutrino and
gravitational wave detectors (such as IceCube and Advanced LIGO),
which may ultimately hold the key to unraveling the mystery of the SN
explosion mechanism, and perhaps the identity of the compact remnants.


\medskip \noindent Correspondence should be addressed to
A.~M.~Soderberg (e-mail: alicia@astro.princeton.edu).

\begin{acknowledge} 
Based in part on observations obtained at the Gemini Observatory 
through the Director's Discretionary Time.  Gemini is operated by 
the Association of Universities for Research in   
Astronomy, Inc., under a cooperative agreement with the NSF on behalf
of the Gemini partnership: the National Science Foundation (United
States), the Science and Technology Facilities Council (United
Kingdom), the National Research Council (Canada), CONICYT (Chile), the
Australian Research Council (Australia), CNPq (Brazil) and SECYT
(Argentina).  The VLA is operated by the National Radio
Astronomy Observatory, a facility of the National Science Foundation
operated under cooperative agreement by Associated Universities, Inc.
Some of the data presented herein were obtained at the
W.~M.~Keck Observatory, which is operated as a scientific partnership
among the California Institute of Technology, the University of
California and the National Aeronautics and Space Administration. The
Observatory was made possible by the generous financial support of the
W.~M.~Keck Foundation.  AMS acknowledges support by NASA through a
Hubble Fellowship grant.  \end{acknowledge}

\clearpage

\noindent {\bf Figure~1:~ Discovery image and X-ray light curve of
XRO\,080109/SN\,2008D} {\it Panel a:} X-ray (left) and UV (right)
images of the field obtained on 2008 Jan 7 UT during \swift\
observations of the Type Ibc SN\,2007uy.  No source is detected at the
position of SN\,2008D to a limit of $\lsim 10^{-3}$ counts s$^{-1}$ in
the X-rays and $U\gsim 20.3$ mag.  {\it Panel b:} Repeated UV and
X-ray observations of the field from Jan 9 UT during which we
serendipitously discovered XRO\,080109 and its UV counterpart.  The
position of XRO\,080109 is $\alpha$=\ra{09}{09}{30.70} and
$\delta$=\dec{33}{08}{19.1} (J2000) ($\pm 3.5$ arcsec), about 9 kpc
from the center of NGC\,2770.  {\it Panel c:} X-ray light curve of
XRO\,080109 in the $0.3-10$ keV band.  The data were accumulated in
the photon counting mode and were processed using version 2.8 of the
\swift\ software package, including the most recent calibration and
exposure maps.  The high count rate resulted in photon pile-up, which
we correct for by fitting a King function profile to the point spread
function (PSF) to determine the radial point at which the measured PSF
deviates from the model.  The counts were extracted using an annular
aperture that excluded the affected 4 pixel core of the PSF, and the
count rate was corrected according to the model.  Using a fast rise,
exponential decay model (red curve) we determine the properties of
the outburst, in particular its onset time, $t_0$, which corresponds to the
explosion time of SN\,2008D.  The best-fit parameters are a peak time
of $63\pm 7$ s after the beginning of the observation, an e-folding
time of $129\pm 6$ s, and peak count rate of $ 6.2\pm 0.4$ counts
s$^{-1}$ (90\% confidence level using Cash statistics).  The best-fit
value of $t_0$ is Jan 9 13:32:40 UT (i.e., 9 s before the start of the
observation) with a 90\% uncertainty range of 13:32:20 to 13:32:48 UT.

\bigskip
\noindent {\bf Figure~2:~ Optical spectra of XRO\,080109/SN\,2008D.}
The observations were performed using the following facilities: The
Gemini Multi-Object Spectrograph (GMOS) on the Gemini-North 8 m
telescope (black); the Dual Imaging Spectrograph (DIS) on the Apache
Point 3.5 m telescope (blue); the Double Spectrograph (DBSP) on the
Palomar Hale 200-inch telescope (green); and the Low Resolution
Spectrograph (LRS) on the Hobby-Eberly 9.2 m telescope (magenta).  The
details of the observational setup and the exposure times are provided
in the Suppl.~Info.  The data were reduced using the {\tt
gemini} package within the Image Reduction and Analysis Facility
(IRAF) software for the GMOS data.  All other observations were
reduced using standard packages in IRAF.  The SN spectra were
extracted from the two-dimensional data using a nearby background
region to reduce the contamination from host galaxy emission.
Absolute flux calibration was achieved using observations of the
standard stars Feige 34 and G191B2B.  The spectra are plotted
logarithmically in flux units and shifted for clarity.  The bottom
panel includes a model fit to the Jan~25 spectrum using the spectral
fitting code SYNOW.  We identify several strong features attributed to
He I, O I, and Fe II indicating a Type Ibc classification.  In
addition, we find an absorption feature at 6200 \AA\ that can be
identified as Si II or high velocity H I (see Suppl.~Info.~for
details).

\bigskip \noindent {\bf Figure~3:~Optical and UV light curves of
XRO\,080109/SN\,2008D.}  Data are from \swift\/UVOT (circles), Palomar
60-inch telescope (P60; squares), Gemini/GMOS (diamonds), and the
SLOTIS telescope (triangles).  Tables summarizing the observations and
data analysis are available in the Suppl.~Info.  The data have not
been corrected for host galaxy extinction and have been offset (as
labeled) for clarity.  We fit the data prior to 3 days with a cooling
envelope blackbody emission model\cite{wmc07} (dashed lines) that
accounts for host extinction ($A_V$).  We find a reasonable fit to the
data with $R_*\approx 10^{11}$ cm, $E_{K}\approx 2\times 10^{51}$ erg,
$M_{\rm ej}\approx 5$ M$_\odot$ and $A_V\approx 1.9$ mag, consistent
with the constraints from the high-resolution optical spectrum.  The
radius and temperature of the photosphere at 1 d are $R_{\rm
ph}\approx 3\times 10^{14}$ cm and $T_{\rm ph}\approx 10^4$ K,
respectively.  The bottom panel shows the absolute bolometric
magnitude light curve (corrected for host extinction), The dashed
lines are the same cooling envelope model described above, while the
dotted lines are models of supernova emission powered by radioactive
decay.  The solid lines are combined models taking into account the
decay of $^{56}$Ni (thin line) and $^{56}$Ni+$^{56}$Co (thick line).
The SN models provide an independent measure of $E_{K}$ and $M_{\rm
ej}$, as well as $M_{\rm Ni}$ (see Suppl.~Info.~for a detailed
discussion of the models).  We find values that are consistent to
within $30\%$ with those inferred from the cooling envelope model.

\bigskip \noindent {\bf Figure~4:~Radio light curves, spectra, and
image of XRO\,080109/SN\,2008D} The radio data from 1.4 to 95 GHz were
obtained with the VLA, CARMA, and the VLBA (circles are detections and
inverted triangles represent $3\sigma$ upper limits).  The flux
measurements and a descripton of the data analysis are provided in the
Supp.~Info.  {\it Panel a:~} Radio light-curves with a model of
synchrotron self-absorbed emission arising\cite{skb+05,chev82} from
shocked material surrounding the freely-expanding SN.  We adopt a
shock compression factor of $\eta=4$ for the post-shock material and
assume that the electrons and magnetic fields each contribute 10\% to
the total post-shock energy density.  The best fit model (solid lines)
implies the following physical parameters and temporal evolution: $R
\approx 3\times 10^{15}(t/5\,{\rm d})^{0.9}$ cm, $E_{K,R}\approx
10^{48} (t/5\,{\rm d})^{0.8}$ erg, and $B\approx 2.4 (t/5{\rm
d})^{-1}$, where $B$ is the magnetic field strength. The implied
density profile is $\rho(r)\propto r^{-2}$, as expected for the wind
from a massive star.  {\it Panel b:} Broadband radio spectra.  The
spectral peak of the radio synchrotron emission cascades to lower
frequencies over the course of our follow-up observations with
$\nu_p\propto t^{-1}$.  The low frequency turn-over is consistent with
expectations for synchrotron self-absorption (grey lines). {\it Panel
c:}~Radio image from a VLBA observation on Feb 8 UT.  We place an
upper limit on the angular size of the ejecta of 1.2 mas ($3\sigma$),
corresponding to a physical radius of $\simlt 2.4\times 10^{17}$ cm.
This limit is factor of 16 times larger, and therefore consistent
with the radius derived from the radio SN model.  However, it places
a limit of $(\gamma\beta)\lsim 3$ on the expansion velocity.

\bigskip 

\noindent {\bf Figure~5:~Volumetric rate of X-ray outbursts similar to
XRO\,080109.}  We use all XRT observations longer than 300 s along
with the field of view (24 arcmin on a side), the number density of
$L_*$ galaxies ($\phi\approx 0.05$ $L_*$ Mpc$^{-3}$), and the
detectability limit of XRT for events like XRO\,080109 ($d\lsim 200$
Mpc).  The curves indicate the rate ($L_*^{-1}$ Mpc$^{-3}$ yr$^{-1}$)
inferred from one detection in a total of about 2 years of effective
on-sky XRT observations as a function of the distance to which XROs
can be detected.  Also shown are the rates\cite{cet99} of
core-collapse SNe (solid horizontal line) and SNe Ibc (dashed
horizontal line) as determined from optical SN searches.  The rate of
events like XRO\,080109 is consistent with the core-collapse rate at
the $50\%$ probability level.

\clearpage

\clearpage

\clearpage

\clearpage

\clearpage

\begin{figure}
\centerline{\psfig{file=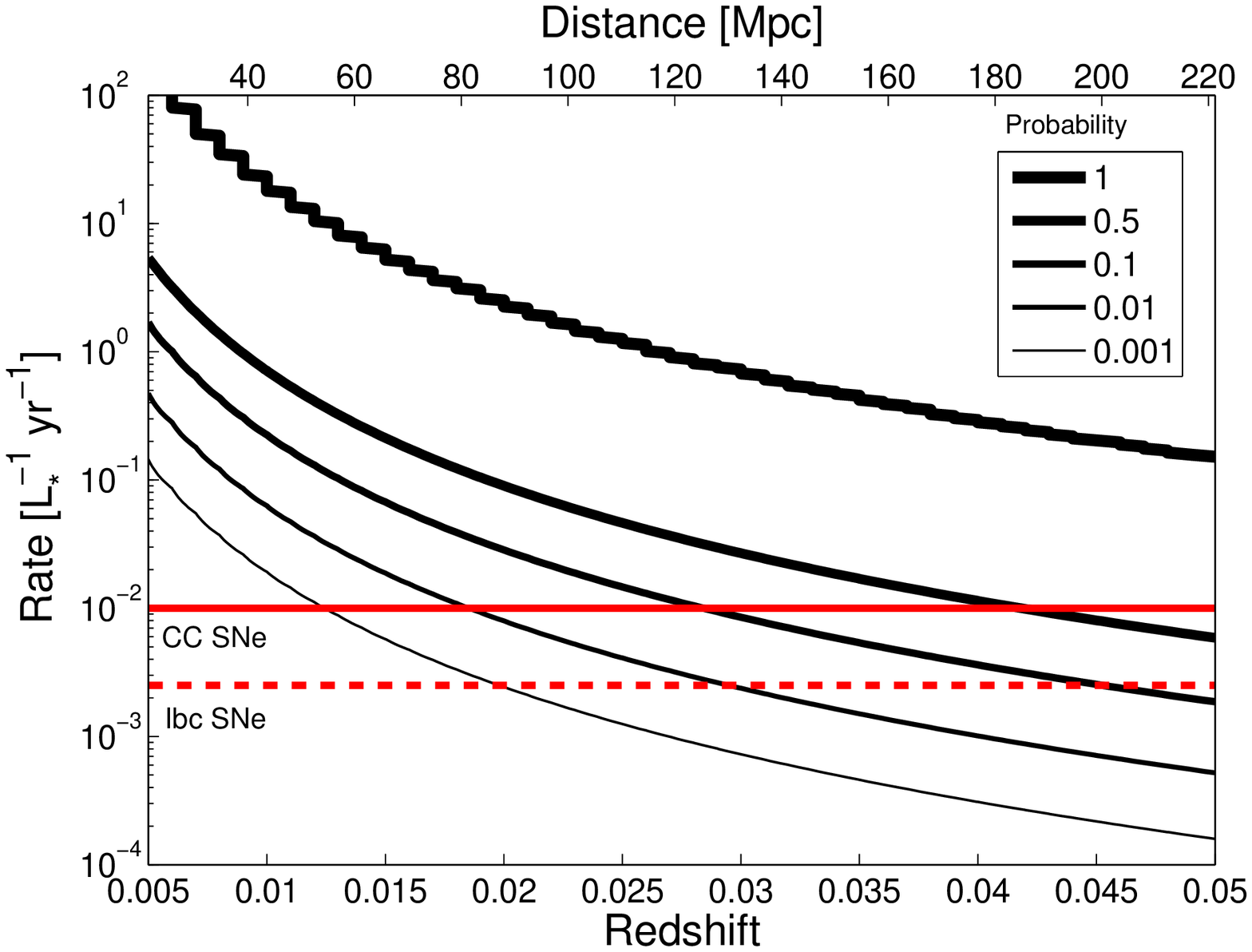,width=6in,angle=0}}
\caption[]{}
\label{fig:rate}
\end{figure}

\end{document}